\documentstyle[prd,aps,preprint,tighten,epsfig]{revtex}

\begin{document}

\draft

\title{A further study of $\mu$-$\tau$ symmetry breaking at
neutrino telescopes after the Daya Bay and RENO measurements of
$\theta^{}_{13}$}
\author{{\bf Zhi-zhong Xing}
\thanks{E-mail: xingzz@mail.ihep.ac.cn}}
\address{Institute of High Energy Physics, Chinese Academy of Sciences,
Beijing 100049, China}

\maketitle

\begin{abstract}
Current neutrino oscillation data indicate that $\theta^{}_{13}$ is
not strongly suppressed and $\theta^{}_{23}$ might have an
appreciable deviation from $\pi/4$, implying that the $3\times 3$
neutrino mixing matrix $V$ does not have an exact $\mu$-$\tau$
permutation symmetry. We make a further study of the effect of
$\mu$-$\tau$ symmetry breaking on the democratic flavor distribution
of ultrahigh-energy (UHE) cosmic neutrinos at a neutrino telescope,
and find that it is characterized by $|V^{}_{\mu i}|^2 - |V^{}_{\tau
i}|^2$ which would vanish if either $\theta^{}_{23} =\pi/4$ and
$\theta^{}_{13} = 0$ or $\theta^{}_{23} = \pi/4$ and $\delta =
\pm\pi/2$ held. We observe that the second-order $\mu$-$\tau$
symmetry breaking term $\overline{\Delta}$ may be numerically
comparable with or even larger than the first-order term $\Delta$ in
the flux ratios $\phi^{\rm T}_e : \phi^{\rm T}_\mu : \phi^{\rm
T}_\tau \simeq (1- 2\Delta) : (1 + \Delta + \overline{\Delta}) : (1
+ \Delta - \overline{\Delta})$, if $\sin (\theta^{}_{23} -\pi/4)$
and $\cos\delta$ have the same sign. The detection of the UHE
$\overline{\nu}^{}_e$ flux via the Glashow-resonance channel
$\overline{\nu}^{}_e e \rightarrow W^- \to {\rm anything}$ is also
discussed by taking account of the first- and second-order
$\mu$-$\tau$ symmetry breaking effects.
\end{abstract}


\newpage

\section{Introduction}

Current experimental data have convinced us that three known
neutrinos can oscillate from one flavor to another \cite{PDG},
implying the existence of a mismatch between their flavor and mass
eigenstates. Hence the lepton flavors must mix as the quark flavors,
and this phenomenon can be described by using an effective $3\times
3$ unitary matrix $V$. In the basis where the flavor and mass
eigenstates of three charged leptons are identical, $V$
provides a unique link between the neutrino flavor $(\nu^{}_e,
\nu^{}_\mu, \nu^{}_\tau)$ and mass $(\nu^{}_1, \nu^{}_2, \nu^{}_3)$
eigenstates:
\begin{eqnarray}
\left ( \matrix{\nu^{}_e \cr \nu^{}_\mu \cr \nu^{}_\tau \cr} \right
) \; =\; \left ( \matrix{ V^{}_{e1} & V^{}_{e2} & V^{}_{e3} \cr
V^{}_{\mu 1} & V^{}_{\mu 2} & V^{}_{\mu 3} \cr V^{}_{\tau 1} &
V^{}_{\tau 2} & V^{}_{\tau 3} \cr} \right ) \left ( \matrix{
\nu^{}_1 \cr \nu^{}_2 \cr \nu^{}_3 \cr} \right ) \; .
\end{eqnarray}
One may parametrize $V$ in terms of three mixing angles and three
phase angles as follows:
\begin{eqnarray}
V = \left( \matrix{ c^{}_{12} c^{}_{13} & s^{}_{12} c^{}_{13} &
s^{}_{13} e^{-i\delta} \cr -s^{}_{12} c^{}_{23} - c^{}_{12}
s^{}_{13} s^{}_{23} e^{i\delta} & c^{}_{12} c^{}_{23} - s^{}_{12}
s^{}_{13} s^{}_{23} e^{i\delta} & c^{}_{13} s^{}_{23} \cr s^{}_{12}
s^{}_{23} - c^{}_{12} s^{}_{13} c^{}_{23} e^{i\delta} & -c^{}_{12}
s^{}_{23} - s^{}_{12} s^{}_{13} c^{}_{23} e^{i\delta} & c^{}_{13}
c^{}_{23} \cr} \right) P^{}_\nu \; ,
\end{eqnarray}
where $c^{}_{ij} \equiv \cos\theta^{}_{ij}$, $s^{}_{ij} \equiv
\sin\theta^{}_{ij}$ (for $ij = 12, 13, 23$), and $P^{}_\nu ={\rm
Diag}\{e^{i\rho}, e^{i\sigma}, 1\}$ is physically relevant only if
massive neutrinos are the Majorana particles. It has been pointed
out that $|V^{}_{\mu i}| = |V^{}_{\tau i}|$ (for $i=1,2,3$) holds
exactly, if either the conditions $\theta^{}_{13} = 0$ and
$\theta^{}_{23} = \pi/4$ or the conditions $\delta = \pm\pi/2$ and
$\theta^{}_{23} = \pi/4$ are satisfied \cite{XZ2008}. However, this
interesting $\mu$-$\tau$ permutation symmetry must be broken: on the
one hand, the fact that $\theta^{}_{13}$ is not strongly suppressed
has recently been established in the Daya Bay and RENO reactor
antineutrino oscillation experiments \cite{DYB}; on the other hand,
the latest global fit of all the available neutrino oscillation data
hints that the value of $\theta^{}_{23}$ might have an appreciable
deviation from $\pi/4$ \cite{Fogli}. An important task in today's
experimental neutrino physics is therefore to determine the strength
of $\mu$-$\tau$ symmetry breaking, so as to help the theorists try
different flavor symmetries and deeply understand the leptonic
flavor mixing structure \cite{Xing2012}.

The neutrino telescopes (e.g., the running IceCube detector at the
South Pole \cite{IC} and the proposed KM3NeT detector in the
Mediterranean Sea \cite{KM3}), which aim  to observe the
ultrahigh-energy (UHE) cosmic neutrinos and their flavor
distributions, can serve as a novel probe of the $\mu$-$\tau$
symmetry breaking effects. It is well known that neutrino
oscillations may map $\phi^{\rm S}_e : \phi^{\rm S}_\mu : \phi^{\rm
S}_\tau = 1 : 2 : 0$, the initial flavor ratios of UHE cosmic
neutrino fluxes produced from $p\gamma$ or $pp$ collisions at a
distant astrophysical source, into $\phi^{\rm T}_e : \phi^{\rm
T}_\mu : \phi^{\rm T}_\tau = 1 : 1 : 1$ at a neutrino telescope on
the Earth \cite{Pakvasa} if there is the exact $\mu$-$\tau$
permutation symmetry. It is also known that such a democratic flavor
distribution can be broken down to \cite{Xing06}
\begin{eqnarray}
\phi^{\rm T}_e : \phi^{\rm T}_\mu : \phi^{\rm T}_\tau \simeq \left(1
-2 \Delta\right) : \left(1 +\Delta\right) : \left(1 +\Delta\right)
\; ,
\end{eqnarray}
where
\begin{eqnarray}
\Delta = \frac{1}{2} \sin^2 2\theta^{}_{12} \sin\varepsilon -
\frac{1}{4} \sin 4\theta^{}_{12} \sin\theta^{}_{13} \cos\delta \; ,
\end{eqnarray}
and $\varepsilon \equiv \theta^{}_{23} -\pi/4$ for the
parametrization of $V$ given in Eq. (2). Hence $\Delta$ signifies
the combined effects of $\mu$-$\tau$ symmetry breaking, and whether
its magnitude can reach the $10\%$ level or not depends crucially on
the sizes and signs of $\sin\varepsilon$ and $\cos\delta$. Note that
the analytical approximation made in Eq. (3) does not reflect the
difference between $\phi^{\rm T}_\mu$ and $\phi^{\rm T}_\tau$, which
should be given by the terms proportional to $\sin^2\varepsilon$,
$\sin^2\theta^{}_{13}$ and $\sin\varepsilon \sin\theta^{}_{13}$
\cite{Rodejohann}.

The purpose of this paper is to make a further study of $\mu$-$\tau$
symmetry breaking relevant to the detection of UHE cosmic neutrinos
originating from a certain cosmic accelerator. First, we calculate
the flavor distribution of UHE cosmic neutrinos at a neutrino
telescope in a parametrization-independent way. Such an exercise
allows us to generalize the approximate result in Eq. (3) to the
following exact one:
\begin{eqnarray}
\phi^{\rm T}_\alpha = \frac{\phi^{}_0}{3} \left[1 + \sum_i
|V^{}_{\alpha i}|^2 \left( |V^{}_{\mu i}|^2 - |V^{}_{\tau i}|^2
\right) \right] \;
\end{eqnarray}
for $\alpha = e$, $\mu$ and $\tau$, where $\phi^{}_0 = \phi^{\rm
S}_e + \phi^{\rm S}_\mu + \phi^{\rm S}_\tau$ is the total flux of
UHE cosmic neutrinos and antineutrinos of all three flavors. Now it
becomes transparent that a deviation from the democratic flavor
distribution $\phi^{\rm T}_e : \phi^{\rm T}_\mu : \phi^{\rm T}_\tau
= 1 : 1 : 1$ at a neutrino telescope is characterized by the
$\mu$-$\tau$ symmetry breaking terms $|V^{}_{\mu i}|^2 - |V^{}_{\tau
i}|^2$. In particular, we find that the difference between
$\phi^{\rm T}_\mu$ and $\phi^{\rm T}_\tau$ is measured by
$(|V^{}_{\mu i}|^2 - |V^{}_{\tau i}|^2)^2$. We also find that the
second-order $\mu$-$\tau$ symmetry breaking term
$\overline{\Delta}$, which is a function of $\sin^2\varepsilon$,
$\sin^2\theta^{}_{13}$ and $\sin\varepsilon \sin\theta^{}_{13}$, may
be numerically comparable with or even larger than $\Delta$ in the
realistic flux ratios $\phi^{\rm T}_e : \phi^{\rm T}_\mu : \phi^{\rm
T}_\tau \simeq (1- 2\Delta) : (1 + \Delta + \overline{\Delta}) : (1
+ \Delta - \overline{\Delta})$ if $\sin\varepsilon$ and $\cos\delta$
have the same sign. Second, we reexamine the effect of $\mu$-$\tau$
symmetry breaking on the $\overline{\nu}^{}_e$ flux of
$E^{}_{\overline{\nu}^{}_e} \approx 6.3 ~ {\rm PeV}$ which is
detectable via the well-known Glashow-resonance (GR) channel
$\overline{\nu}^{}_e e \rightarrow W^- \rightarrow ~ {\rm anything}$
\cite{Glashow} at neutrino telescopes. Different from the previous
result obtained Ref. \cite{Xing06}, our new result for the
GR-mediated $\overline{\nu}^{}_e$ events will include the
second-order $\mu$-$\tau$ symmetry breaking effect.

\section{Effects of $\mu$-$\tau$ symmetry breaking}

Let us define $\phi^{\rm S}_\alpha \equiv \phi^{\rm
S}_{\nu^{}_\alpha} + \phi^{\rm S}_{\overline{\nu}^{}_\alpha}$ and
$\phi^{\rm T}_\alpha \equiv \phi^{\rm T}_{\nu^{}_\alpha} + \phi^{\rm
T}_{\overline{\nu}^{}_\alpha}$ (for $\alpha = e, \mu, \tau$)
throughout this paper, where $\phi^{\rm S}_{\nu^{}_\alpha}$ (or
$\phi^{\rm T}_{\nu^{}_\alpha}$) and $\phi^{\rm
S}_{\overline{\nu}^{}_\alpha}$ (or $\phi^{\rm
T}_{\overline{\nu}^{}_\alpha}$) denote the $\nu^{}_\alpha$ and
$\overline{\nu}^{}_\alpha$ fluxes at a distant astrophysical source
(or at a neutrino telescope), respectively. For most of the
currently-envisaged sources of UHE cosmic neutrinos \cite{XZ2011}, a
general expectation is that the initial neutrino fluxes are produced
via the decay chain of charged pions and muons created from $pp$ or
$p\gamma$ collisions and their flavor content can be expressed as
\begin{eqnarray}
\left \{\phi^{\rm S}_e ~,~ \phi^{\rm S}_\mu ~,~ \phi^{\rm S}_\tau
\right \} \; = \; \left \{ \frac{1}{3} ~,~ \frac{2}{3} ~,~ 0 \right
\} \phi^{}_0 \; ,
\end{eqnarray}
where $\phi^{S}_\tau = \phi^{S}_{\nu^{}_\tau} =
\phi^{S}_{\overline{\nu}^{}_\tau} = 0$, and $\phi^{}_0 = \phi^{\rm
S}_e + \phi^{\rm S}_\mu + \phi^{\rm S}_\tau$ is the total flux of
neutrinos and antineutrinos of all flavors. Thanks to neutrino
oscillations, the flavor distribution of such UHE cosmic neutrinos
at a neutrino telescope is described by
\begin{eqnarray}
\phi^{\rm T}_\alpha = \phi^{\rm T}_{\nu^{}_\alpha} + \phi^{\rm
T}_{\overline{\nu}^{}_\alpha} = \sum_\beta \left[\phi^{\rm
S}_{\nu^{}_\beta} P(\nu^{}_\beta \to \nu^{}_\alpha) + \phi^{\rm
S}_{\overline{\nu}^{}_\beta} P(\overline{\nu}^{}_\beta \to
\overline{\nu}^{}_\alpha) \right] = \sum_i \sum_\beta
\left(|V^{}_{\alpha i}|^2 |V^{}_{\beta i}|^2 \phi^{\rm S}_\beta
\right) \; ,
\end{eqnarray}
where we have used
\begin{eqnarray}
P(\nu^{}_\beta \to \nu^{}_\alpha) = P(\overline{\nu}^{}_\beta \to
\overline{\nu}^{}_\alpha) = \sum_i |V^{}_{\alpha i}|^2 |V^{}_{\beta
i}|^2 \; .
\end{eqnarray}
Since the Galactic distance that the UHE cosmic neutrinos travel far
exceeds the observed neutrino oscillation lengths, $P(\nu^{}_\beta
\to \nu^{}_\alpha)$ and $P(\overline{\nu}^{}_\beta \to
\overline{\nu}^{}_\alpha)$ are actually averaged over many
oscillations and thus become energy- and distance-independent.
Combining Eq. (7) with Eq. (6) and using the unitarity conditions of
$V$, we find
\begin{eqnarray}
\phi^{\rm T}_\alpha & = & \frac{\phi^{}_0}{3} \sum_i |V^{}_{\alpha
i}|^2 \left(|V^{}_{e i}|^2 + 2 |V^{}_{\mu i}|^2 \right) =
\frac{\phi^{}_0}{3} \sum_i |V^{}_{\alpha i}|^2 \left(1 + |V^{}_{\mu
i}|^2 - |V^{}_{\tau i}|^2 \right) \nonumber \\
& = & \frac{\phi^{}_0}{3} \left[1 + \sum_i |V^{}_{\alpha
i}|^2 \left( |V^{}_{\mu i}|^2 - |V^{}_{\tau i}|^2 \right) \right] \; .
\end{eqnarray}
This is a simple proof of Eq. (5). Of course, the relationship
$\phi^{\rm T}_e + \phi^{\rm T}_\mu + \phi^{\rm T}_\tau = \phi^{}_0$
holds. To be more explicit, we have
\begin{eqnarray}
\phi^{\rm T}_e & = & \frac{\phi^{}_0}{3} \left[1 + \sum_i |V^{}_{e
i}|^2 \left( |V^{}_{\mu i}|^2 - |V^{}_{\tau i}|^2 \right) \right] \;
, \nonumber \\
\phi^{\rm T}_\mu & = & \frac{\phi^{}_0}{3} \left[1 + \sum_i |V^{}_{\mu
i}|^2 \left( |V^{}_{\mu i}|^2 - |V^{}_{\tau i}|^2 \right) \right] \; ,
\nonumber \\
\phi^{\rm T}_\tau & = & \frac{\phi^{}_0}{3} \left[1 + \sum_i |V^{}_{\tau
i}|^2 \left( |V^{}_{\mu i}|^2 - |V^{}_{\tau i}|^2 \right) \right] \; ,
\end{eqnarray}
from which we obtain the difference between $\phi^{\rm T}_\mu$ and
$\phi^{\rm T}_\tau$ as
\begin{eqnarray}
\phi^{\rm T}_\mu - \phi^{\rm T}_\tau = \frac{\phi^{}_0}{3}
\sum_i \left( |V^{}_{\mu i}|^2 - |V^{}_{\tau i}|^2 \right)^2 \; .
\end{eqnarray}
Now it becomes quite transparent that the deviation of $\phi^{\rm
T}_\alpha$ (for $\alpha = e, \mu, \tau$) from $\phi^{}_0/3$ is
measured by $|V^{}_{\mu i}|^2 - |V^{}_{\tau i}|^2$, and the
difference between $\phi^{\rm T}_\mu$ and $\phi^{\rm T}_\tau$ is
purely an effect governed by $(|V^{}_{\mu i}|^2 - |V^{}_{\tau
i}|^2)^2$. This exact and parametrization-independent observation is
therefore useful for us to probe the leptonic flavor mixing
structure via the detection of UHE cosmic neutrinos at neutrino
telescopes.

Considering that the $\mu$-$\tau$ symmetry of $V$ is
possible to be partly or softly broken, we define the following
three $\mu$-$\tau$ symmetry breaking quantities and express them in
terms of three neutrino mixing angles and the Dirac CP-violating
phase in the standard parametrization of $V$ as given in Eq. (2):
\begin{eqnarray}
\Delta^{}_1 & \equiv & |V^{}_{\mu 1}|^2 - |V^{}_{\tau 1}|^2 = \left(
\sin^2\theta^{}_{12} - \cos^2\theta^{}_{12} \sin^2\theta^{}_{13}
\right) \cos 2\theta^{}_{23} + \sin 2\theta^{}_{12} \sin
2\theta^{}_{23} \sin\theta^{}_{13} \cos\delta \; , ~~ \nonumber \\
\Delta^{}_2 & \equiv & |V^{}_{\mu 2}|^2 - |V^{}_{\tau 2}|^2 = \left(
\cos^2\theta^{}_{12} - \sin^2\theta^{}_{12} \sin^2\theta^{}_{13}
\right) \cos 2\theta^{}_{23} - \sin 2\theta^{}_{12} \sin
2\theta^{}_{23} \sin\theta^{}_{13} \cos\delta \; , \nonumber \\
\Delta^{}_3 & \equiv & |V^{}_{\mu 3}|^2 - |V^{}_{\tau 3}|^2 =
-\cos^2\theta^{}_{13} \cos 2\theta^{}_{23} \; .
\end{eqnarray}
We see that the relationship $\Delta^{}_1 + \Delta^{}_2 +
\Delta^{}_3 = 0$ holds exactly, as guaranteed by the unitarity of
$V$. There are three special but interesting cases, in which the
$\mu$-$\tau$ permutation symmetry is not completely broken:
\begin{itemize}

\item     Case (A): $\theta^{}_{23} \to \pi/4$. In this limit,
$\Delta^{}_i$ (for $i=1,2,3$) can be simplified to
\begin{eqnarray}
\Delta^{}_1 & = & -\Delta^{}_2 = \sin 2\theta^{}_{12}
\sin\theta^{}_{13} \cos\delta \; , ~~~~~~~~~~ \;\;
\nonumber \\
\Delta^{}_3 & = & 0 \; .
\end{eqnarray}

\item     Case (B): $\theta^{}_{13} \to 0$. In this limit, the results
in Eq. (12) are simplified to
\begin{eqnarray}
\Delta^{}_1 & = & \tan^2 \theta^{}_{12} \Delta^{}_2 = \sin^2
\theta^{}_{12} \cos 2\theta^{}_{23} \; , ~~~~~~~~~
\nonumber \\
\Delta^{}_3 & = & - \cos 2\theta^{}_{23} \; .
\end{eqnarray}

\item     Case (C): $\delta \to \pm \pi/2$. In this limit, we simply
obtain
\begin{eqnarray}
\Delta^{}_1 & = & \left( \sin^2\theta^{}_{12} - \cos^2\theta^{}_{12}
\sin^2\theta^{}_{13}
\right) \cos 2\theta^{}_{23} \; ,  ~ \nonumber \\
\Delta^{}_2 & = & \left( \cos^2\theta^{}_{12} - \sin^2\theta^{}_{12}
\sin^2\theta^{}_{13}
\right) \cos 2\theta^{}_{23} \; , \nonumber \\
\Delta^{}_3 & = & -\cos^2\theta^{}_{13} \cos 2\theta^{}_{23} \; .
\end{eqnarray}
\end{itemize}
Current experimental data tell us that case (A) is probably not
true, case (B) is definitely not true, and case (C) remains an
interesting possibility. In any case, we may make analytical
approximations for $\phi^{\rm T}_\alpha$ (for $\alpha =e, \mu,
\tau$) up to the accuracy of $\sin^2\varepsilon$,
$\sin^2\theta^{}_{13}$ and $\sin\varepsilon \sin\theta^{}_{13}$.
Then a result similar to Eq. (3) is
\begin{eqnarray}
\phi^{\rm T}_e : \phi^{\rm T}_\mu : \phi^{\rm T}_\tau
\simeq (1- 2\Delta) : (1 + \Delta + \overline{\Delta}) :
(1 + \Delta - \overline{\Delta}) \; ,
\end{eqnarray}
where $\Delta$ has been given in Eq. (4), and $\overline{\Delta}$ is
defined as
\footnote{This second-order perturbation term is consistent with the
one obtained in Ref. \cite{Rodejohann}, where the definition of
$\varepsilon$ has an opposite sign.}
\begin{eqnarray}
\overline{\Delta} = \left(4 - \sin^2 2\theta^{}_{12} \right)
\sin^2\varepsilon + \sin^2 2\theta^{}_{12}
\sin^2\theta^{}_{13}\cos^2\delta + \sin 4\theta^{}_{12}
\sin\varepsilon \sin\theta^{}_{13} \cos\delta \; .
\end{eqnarray}
It is easy to show that $\overline{\Delta} \geq 0$ holds for
arbitrary values of $\delta$, because the above expression can
simply be transformed into $\overline{\Delta} = 3 \sin^2\varepsilon
+ \left( \cos 2\theta^{}_{12} \sin\varepsilon + \sin 2\theta^{}_{12}
\sin\theta^{}_{13} \cos\delta \right)^2$, which may vanish if both
$\sin\varepsilon =0$ and $\sin\theta^{}_{13} =0$ (or $\cos\delta =
0$) hold.

One may define three working observables at neutrino telescopes
\cite{XZ} and link them to the $\mu$-$\tau$ symmetry breaking
quantities:
\begin{eqnarray}
R^{}_e & \equiv & \frac{\phi^{\rm T}_e}{\phi^{\rm T}_\mu + \phi^{\rm
T}_\tau} = \frac{1 + \displaystyle\sum_i |V^{}_{e i}|^2
\Delta^{}_i}{2 - \displaystyle\sum_i |V^{}_{e i}|^2 \Delta^{}_i}
\simeq \frac{1}{2} \left(1+ \frac{3}{2} \sum_i |V^{}_{e i}|^2
\Delta^{}_i \right)
\simeq \frac{1}{2} - \frac{3}{2} \Delta \; , \nonumber \\
R^{}_\mu & \equiv & \frac{\phi^{\rm T}_\mu}{\phi^{\rm T}_\tau +
\phi^{\rm T}_e} = \frac{1 + \displaystyle\sum_i |V^{}_{\mu i}|^2
\Delta^{}_i}{2 - \displaystyle\sum_i |V^{}_{\mu i}|^2 \Delta^{}_i}
\simeq \frac{1}{2} \left(1+ \frac{3}{2} \sum_i |V^{}_{\mu i}|^2
\Delta^{}_i \right) \simeq \frac{1}{2} + \frac{3}{4} \left(\Delta +
\overline{\Delta}
\right) \; , \nonumber \\
R^{}_\tau & \equiv & \frac{\phi^{\rm T}_\tau}{\phi^{\rm T}_e +
\phi^{\rm T}_\mu} = \frac{1 + \displaystyle\sum_i |V^{}_{\tau i}|^2
\Delta^{}_i}{2 - \displaystyle\sum_i |V^{}_{\tau i}|^2 \Delta^{}_i}
\simeq \frac{1}{2} \left(1+ \frac{3}{2} \sum_i |V^{}_{\tau i}|^2
\Delta^{}_i \right) \simeq \frac{1}{2} + \frac{3}{4} \left(\Delta -
\overline{\Delta} \right) \; ,
\end{eqnarray}
Comparing Eq. (11) and Eqs. (16) --- (18), we immediately obtain
\begin{eqnarray}
\phi^{\rm T}_\mu - \phi^{\rm T}_\tau & = & \frac{\phi^{}_0}{3}
\sum_i \Delta^2_i \simeq \frac{2\phi^{}_0}{3} \overline{\Delta} \; ,
\nonumber \\
R^{}_\mu - R^{}_\tau & \simeq & \frac{3}{4} \sum_i \Delta^2_i \simeq
\frac{3}{2} \overline{\Delta} \; .
\end{eqnarray}
Hence $\overline{\Delta}$ signifies the second-order effect of
$\mu$-$\tau$ symmetry breaking, and the departure of $R^{}_\alpha$
(for $\alpha =e, \mu, \tau$) from $1/2$ is a clear measure of the
overall $\mu$-$\tau$ symmetry breaking effects.

To illustrate the possible size of $\mu$-$\tau$ symmetry breaking,
we estimate $\Delta^{}_i$ (for $i=1,2,3$), $\Delta$ and
$\overline{\Delta}$ by using the latest values of $\theta^{}_{12}$,
$\theta^{}_{13}$, $\theta^{}_{23}$ and $\delta$ obtained from a
global analysis of the presently available neutrino oscillation data
done by Fogli {\it et al} \cite{Fogli}. Our numerical results are
listed in TABLE I, where both normal and inverted neutrino mass
hierarchies are taken into account in accordance with Ref.
\cite{Fogli}. Three comments are in order.
\begin{itemize}
\item     $|\Delta^{}_i| \sim 0.1$ (for $i=1,2,3$) holds when
the best-fit values of $\theta^{}_{12}$, $\theta^{}_{13}$,
$\theta^{}_{23}$ and $\delta$ are used. Given the $2\sigma$
intervals of the four input parameters, $|\Delta^{}_i| \sim 0$ is
allowed except for $\Delta^{}_3$ in the case of a normal neutrino
mass hierarchy. It is therefore desirable to determine the departure
of $\theta^{}_{23}$ from $\pi/4$ and that of $\delta$ from
$\pm\pi/2$ in the future experiments.

\item     Since the best-fit values of $\theta^{}_{23}$ and
$\delta$ lie in the ranges $0 < \theta^{}_{23} < \pi/4$ and
$\pi/2 < \delta < \pi$ respectively, $\sin\varepsilon$ and
$\cos\delta$ have the same sign and thus the two terms of
$\Delta$ in Eq. (3) significantly cancel each other. This
significant cancellation leads to $|\Delta| < \overline{\Delta}$
at the percent level, implying a fairly good flavor democracy
for $\phi^{\rm T}_e$, $\phi^{\rm T}_\mu$ and $\phi^{\rm T}_\tau$.

\item     When the $2\sigma$ intervals of $\theta^{}_{12}$, $\theta^{}_{13}$,
$\theta^{}_{23}$ and $\delta$ are taken into account, the lower
bound of $\Delta$ and the upper bound of $\overline{\Delta}$ are
about $0.1$ in magnitude. Hence the flavor democracy of $\phi^{\rm
T}_e$, $\phi^{\rm T}_\mu$ and $\phi^{\rm T}_\tau$ can maximally be
broken at the same level. Needless to say, an appreciable departure
of $R^{}_\alpha$ (for $\alpha=e,\mu,\tau$) from $1/2$ requires an
appreciable effect of $\mu$-$\tau$ symmetry breaking.
\end{itemize}
The rapid development of reactor antineutrino oscillation
experiments implies that $\theta^{}_{13}$ will soon become the best
known angle of $V$. In this case, improving the precision of
$\theta^{}_{23}$ and determining the value of $\delta$ turn out to
be two burning issues which will allow us to pin down the leptonic
flavor mixing structure including CP violation.

\section{On the Glashow resonance}

Let us proceed to look at the effect of $\mu$-$\tau$ symmetry
breaking on the flavor distribution of UHE cosmic neutrinos at a
neutrino telescope by detecting the $\overline{\nu}^{}_e$ flux from
a distant astrophysical source through the GR channel
$\overline{\nu}^{}_e e \rightarrow W^- \rightarrow ~ {\rm anything}$
\cite{Glashow}. This reaction can take place over a narrow energy
interval around the $\overline{\nu}^{}_e$ energy $E^{\rm
GR}_{\overline{\nu}^{}_e} \approx M^2_W/2m^{}_e \approx 6.3 ~ {\rm
PeV}$, and its cross section is about two orders of magnitude larger
than the cross sections of $\overline{\nu}^{}_e N$ interactions of
the same $\overline{\nu}^{}_e$ energy \cite{Gandhi}. A measurement
of the GR reaction is important in neutrino astronomy because it may
serve as a sensitive discriminator of UHE cosmic neutrinos
originating from $p\gamma$ and $pp$ collisions
\cite{G1,G2,XZ11,Lin}. We hope that a neutrino telescope may measure
both the GR-mediated $\overline{\nu}^{}_e$ events and the
$\nu^{}_\mu + \overline{\nu}^{}_\mu$ events of charged-current
interactions in the vicinity of $E^{\rm GR}_{\overline{\nu}^{}_e}$,
and their ratio can be related to the ratio of the
$\overline{\nu}^{}_e$ flux to the $\nu^{}_\mu$ and
$\overline{\nu}^{}_\mu$ fluxes entering the detector:
\begin{eqnarray}
R^{}_{\rm GR} \; \equiv \; \frac{\phi^{\rm
T}_{\overline{\nu}^{}_e}}{\phi^{\rm T}_{\nu^{}_\mu} + \phi^{\rm
T}_{\overline{\nu}^{}_\mu}} = \frac{\phi^{\rm
T}_{\overline{\nu}^{}_e}}{\phi^{\rm T}_\mu} \; .
\end{eqnarray}
Here we follow Ref. \cite{Xing06} to reexamine the $\mu$-$\tau$
symmetry breaking effect on $R^{}_{\rm GR}$.

The initial UHE cosmic neutrino fluxes are produced via the decay
chain of charged pions and muons created from $pp$ or $p\gamma$
collisions at a cosmic accelerator, and thus their flavor
distribution can be expressed as
\begin{eqnarray}
\left \{\phi^{\rm S}_{\nu^{}_e} ,~ \phi^{\rm
S}_{\overline{\nu}^{}_e} ,~ \phi^{\rm S}_{\nu^{}_\mu} ,~ \phi^{\rm
S}_{\overline{\nu}^{}_\mu} ,~ \phi^{\rm S}_{\nu^{}_\tau} ,~
\phi^{\rm S}_{\overline{\nu}^{}_\tau} \right \} = \left\{
\begin{array}{lcl} \displaystyle \left \{ \frac{1}{6} ~,~
\frac{1}{6} ~,~ \frac{1}{3} ~,~ \frac{1}{3} ~,~ 0 ~,~ 0 \right \}
\phi^{}_0 ~ && ~ (pp ~ {\rm collisions}) \; , \\ \vspace{-.2cm} \\
\displaystyle \left \{ \frac{1}{3} ~,~ 0 ~,~ \frac{1}{3} ~,~
\frac{1}{3} ~,~ 0 ~,~ 0 \right \} \phi^{}_0 ~ && ~ (p\gamma ~ {\rm
collisions}) \; .
\end{array} \right .
\end{eqnarray}
In either case the sum of $\phi^{\rm S}_{\nu^{}_\alpha}$ and
$\phi^{\rm S}_{\overline{\nu}^{}_\alpha}$ (for $\alpha = e, \mu,
\tau$) is consistent with $\phi^{\rm S}_\alpha$ in Eq. (6). Thanks
to neutrino oscillations, the $\overline{\nu}^{}_e$ flux at a
neutrino telescope is given by
\begin{eqnarray}
\phi^{\rm T}_{\overline{\nu}^{}_e} = \sum_\beta \left[\phi^{\rm
S}_{\overline{\nu}^{}_\beta} P(\overline{\nu}^{}_\beta \to
\overline{\nu}^{}_e) \right] = \sum_i \sum_\beta \left(|V^{}_{e
i}|^2 |V^{}_{\beta i}|^2 \phi^{\rm S}_{\overline{\nu}^{}_\beta}
\right) \; .
\end{eqnarray}
To be explicit,
\begin{eqnarray}
\phi^{\rm T}_{\overline{\nu}^{}_e} (pp) & = & \frac{\phi^{}_0}{6}
\left(1 + \sum_i |V^{}_{e i}|^2 \Delta^{}_i \right) \simeq
\frac{\phi^{}_0}{6} \left (1 - 2 \Delta \right ) \; ,
\nonumber \\
\phi^{\rm T}_{\overline{\nu}^{}_e} (p\gamma) & = &
\frac{\phi^{}_0}{3} \sum_i |V^{}_{e i}|^2 |V^{}_{\mu i}|^2 \simeq
\frac{\phi^{}_0}{12} \left [\sin^2 2\theta^{}_{12} - 4 \Delta + 2
\left(1 + \cos^2 2\theta^{}_{12} \right ) \sin^2\theta^{}_{13}
\right] \; .
\end{eqnarray}
Since the expression of $\phi^{\rm T}_\mu$ can be found in Eq. (10),
it is straightforward to calculate $R^{}_{\rm GR}$ by using Eqs.
(20) and (23) for two different astrophysical sources:
\begin{eqnarray}
R^{}_{\rm GR} (pp) & \simeq & \frac{1}{2} - \frac{3}{2} \Delta - \frac{1}{2}
\overline{\Delta} \; ,
\nonumber \\
R^{}_{\rm GR} (p\gamma) & \simeq & \frac{\sin^2 2\theta^{}_{12}}{4}
- \frac{4 + \sin^2 2\theta^{}_{12}}{4} \Delta -
\frac{\sin^2 2\theta^{}_{12}}{4} \overline{\Delta} +
\frac{1+\cos^2 2\theta^{}_{12}}{2} \sin^2 \theta^{}_{13} \; .
\end{eqnarray}
We see that the deviation of $R^{}_{\rm GR} (pp)$ from $1/2$ and
that of $R^{}_{\rm GR} (p\gamma)$ from $\sin^2 2\theta^{}_{12}/4$
are both controlled by the effects of $\mu$-$\tau$ symmetry
breaking, which can maximally be of ${\cal O}(0.1)$. As discussed in
Refs. \cite{G1} and \cite{XZ11}, the IceCube detector running at the
South Pole has a good discovery potential to measure $R^{}_{\rm GR}
(pp)$ after several years of data accumulation. In comparison, it
seems more difficult to probe the GR-mediated UHE
$\overline{\nu}^{}_e$ events originating from the pure $p\gamma$
collisions at a cosmic accelerator.

Of course, one may also consider some other possible astrophysical
sources of UHE cosmic neutrinos, such as the neutron beam source
\cite{Neutron} with $\{\phi^{\rm S}_e : \phi^{\rm S}_\mu : \phi^{\rm
S}_\tau\} = \{1 : 0 : 0\}$ and the muon-damped source \cite{Muon}
with $\{\phi^{\rm S}_e : \phi^{\rm S}_\mu : \phi^{\rm S}_\tau\} =
\{0 : 1 : 0\}$, to study their flavor distributions at neutrino
telescopes and probe the effects of $\mu$-$\tau$ symmetry breaking.

\section{Summary and further discussions}

With the development of several neutrino telescope experiments, a
lot of interest has recently been paid to the flavor issues of UHE
cosmic neutrino fluxes and whether they are detectable in the
foreseeable future \cite{Flavor}. In view of the fact that the
smallest neutrino mixing angle $\theta^{}_{13}$ is not strongly
suppressed \cite{DYB} and the hint that the largest neutrino mixing
angle $\theta^{}_{23}$ might have an appreciable departure from
$\pi/4$ \cite{Fogli}, we have carried out a further study of the
effect of $\mu$-$\tau$ symmetry breaking on the presumably
democratic flavor distribution of UHE cosmic neutrinos at a neutrino
telescope. Our new results are different from the previous ones in
the following aspects:
\begin{itemize}
\item     An exact and parametrization-independent expression for
the fluxes of UHE cosmic neutrinos at a neutrino telescope is
obtained in Eq. (5) or Eq. (10), and it clearly shows the
$\mu$-$\tau$ symmetry breaking effect measured by $|V^{}_{\mu i}|^2
- |V^{}_{\tau i}|^2$ (for $i=1,2,3$). In addition, the difference
between $\phi^{\rm T}_\mu$ and $\phi^{\rm T}_\tau$ is the pure
second-order $\mu$-$\tau$ symmetry breaking effect proportional to
$(|V^{}_{\mu i}|^2 - |V^{}_{\tau i}|^2)^2$.

\item     The first- and second-order $\mu$-$\tau$ symmetry breaking
effects, characterized respectively by $\Delta$ and
$\overline{\Delta}$ in the flux ratios $\phi^{\rm T}_e : \phi^{\rm
T}_\mu : \phi^{\rm T}_\tau \simeq (1- 2\Delta) : (1 + \Delta +
\overline{\Delta}) : (1 + \Delta - \overline{\Delta})$, may be
numerically at the same order of magnitude. This observation is
particularly true when $\sin\varepsilon$ and $\cos\delta$ have the
same sign such that the two first-order $\mu$-$\tau$ symmetry
breaking terms of $\Delta$ cancel each other and then lead us to
$|\Delta| \lesssim \overline{\Delta}$.

\item     $\overline{\Delta} \geq 0$ holds for arbitrary values of $\delta$,
and its contributions to the flux ratios $R^{}_{\rm GR}(pp)$ and
$R^{}_{\rm GR}(p\gamma)$ for the GR-mediated $\overline{\nu}^{}_e$
events are taken into account in Eq. (24). The term proportional to
$\sin^2\theta^{}_{13}$ is also included in the expression of
$R^{}_{\rm GR}(p\gamma)$, but such a term does not appear in
$R^{}_{\rm GR}(pp)$.
\end{itemize}
Because of the poor data on $\varepsilon$ and $\delta$, whether the
magnitude of $\Delta$ and $\overline{\Delta}$ (or one of them) can
be as large as about $10\%$ remains an open question. It is also
possible that both of them are at the $1\%$ level, as shown in TABLE
I with the present best-fit results of $\theta^{}_{23}$ and $\delta$
\cite{Fogli}. If $|\Delta|$ and $\overline{\Delta}$ are finally
confirmed to be really small, then an approximately democratic
flavor distribution $\phi^{\rm T}_e : \phi^{\rm T}_\mu : \phi^{\rm
T}_\tau \simeq 1 : 1 : 1$ will show up at neutrino telescopes for
the UHE cosmic neutrinos originating from $pp$ and (or) $p\gamma$
collisions at a distant astrophysical source.

So far we have omitted the uncertainties associated 
with the initial neutrino fluxes at a given astrophysical source. 
A more careful analysis of the flavor ratios of
UHE cosmic neutrinos originating from $pp$ or $p\gamma$ collisions 
yields $\phi^{\rm S}_e : \phi^{\rm S}_\mu : \phi^{\rm
S}_\tau \simeq 1 : 2\left(1 -\eta\right) : 0$ with $\eta \simeq 0.08$
\cite{eta}. If this uncertainty is taken into account, Eq. (5) or
Eq. (9) will change to
\begin{eqnarray}
\phi^{\rm T}_\alpha \simeq \frac{\phi^{}_0}{3} \left\{1 + \sum_i
|V^{}_{\alpha i}|^2 \left[ \left(|V^{}_{\mu i}|^2 - |V^{}_{\tau i}|^2 
\right) - 2\eta |V^{}_{\mu i}|^2 \right] \right\} \;
\end{eqnarray}
for $\alpha = e$, $\mu$ and $\tau$. This result clearly shows that
the $\eta$-induced correction is in general comparable with (or
even larger than) the effect of $\mu$-$\tau$ symmetry breaking. On
the other hand, there exist the uncertainties associated with the 
identification of different flavors at the neutrino telescope. Taking
the IceCube detector as an example, Beacom {\it et al} \cite{Beacom}
have pointed out that the relative experimental error for determining 
the ratio of the muon track to the non-muon shower is typically 
$\xi \sim 20\%$, depending on the event numbers. This estimate means
that the ratio $R^{}_\mu$ in Eq. (18) may in practice be contaminated 
by $\xi$, which is very likely to overwhelm the $\mu$-$\tau$ symmetry
breaking effect $(\Delta + \overline{\Delta})$. 

We also admit that our treatment does not take account of the other
complexities and uncertainties associated with the origin of UHE
cosmic neutrinos, such as their energy dependence, the effect of
magnetic fields and even possible new physics \cite{Winter}. It is
still too early to say that we have correctly understood the
production mechanism of UHE cosmic rays and neutrinos from a given
cosmic accelerator. But the progress made in the measurement of
lepton flavor mixing parameters is quite encouraging, and it may
finally allow us to well control the error bars from particle
physics (e.g., the effect of $\mu$-$\tau$ symmetry breaking) and
thus concentrate on the unknowns from astrophysics (e.g., the
initial flavor composition of UHE cosmic neutrinos). We believe that
any constraint on the flavor distribution of UHE cosmic neutrinos to
be achieved from a neutrino telescope will be greatly useful in
diagnosing the astrophysical sources and in understanding the
properties of neutrinos themselves. Much more efforts are therefore
needed to make in this direction.

\acknowledgments{The author would like to thank G.L. Fogli and E.
Lisi for stimulating communications about possible implications of
their global-fit results. He is also grateful to Y.F. Li, S. Luo and
Y.L. Zhou for useful discussions or technical helps. This work was
supported in part by the National Natural Science Foundation of
China under Grant No. 11135009.}

\newpage

\begin{table}
\caption{Possible sizes of $\mu$-$\tau$ symmetry breaking at
neutrino telescopes. The values of $\theta^{}_{12}$,
$\theta^{}_{13}$, $\theta^{}_{23}$ and $\delta$ are taken from the
latest global analysis of currently available neutrino oscillation
data done by Fogli {\it et al} [4], where both normal and inverted
neutrino mass hierarchies are considered.}
\begin{center}
\begin{tabular}{lllll}
& Normal hierarchy & Best fit & 2$\sigma$
range \\ \hline
& $\sin^2 \theta^{}_{12}$ & $3.07 \times 10^{-1}$ & $\left(2.75 \cdots
3.42\right) \times 10^{-1}$ \\
& $\sin^2 \theta^{}_{13}$ & $2.45 \times 10^{-2}$ & $\left(1.81 \cdots
3.11\right) \times 10^{-2}$ \\
& $\sin^2 \theta^{}_{23}$ & $3.98 \times 10^{-1}$ & $\left(3.50 \cdots
4.75\right) \times 10^{-1}$ \\
& $\delta$ & $0.89 \times \pi$ & $0 \cdots 2\pi$ \\ \hline
& $\Delta^{}_{1}$ & $-7.38 \times 10^{-2}$ & $-0.151 \cdots +0.256$ \\
& $\Delta^{}_{2}$ & $+2.73 \times 10^{-1}$ & $-0.135 \cdots +0.365$ \\
& $\Delta^{}_{3}$ & $-1.99 \times 10^{-1}$ & $-0.295 \cdots -0.048$ \\
& $\Delta$ & $-1.74 \times 10^{-2}$ & $-0.096 \cdots +0.026$ \\
& $\overline{\Delta}$ & $+6.23 \times 10^{-2}$ &
$0 \cdots 0.120$ \\ \hline\hline
& Inverted hierarchy & Best fit & 2$\sigma$ range
\\ \hline
& $\sin^2 \theta^{}_{12}$ & $3.07 \times 10^{-1}$ & $\left(2.75 \cdots
3.42\right) \times 10^{-1}$ \\
& $\sin^2 \theta^{}_{13}$ & $2.46 \times 10^{-2}$ & $\left(1.83 \cdots
3.13\right) \times 10^{-2}$ \\
& $\sin^2 \theta^{}_{23}$ & $4.08 \times 10^{-1}$ & $\left(3.55 \cdots
6.27\right) \times 10^{-1}$ \\
& $\delta$ & $0.90 \times \pi$ & $0 \cdots 2\pi$ \\ \hline
& $\Delta^{}_{1}$ & $-8.19 \times 10^{-2}$ & $-0.244 \cdots +0.254$ \\
& $\Delta^{}_{2}$ & $+2.61 \times 10^{-1}$ & $-0.335 \cdots +0.359$ \\
& $\Delta^{}_{3}$ & $-1.79 \times 10^{-1}$ & $-0.285 \cdots +0.249$ \\
& $\Delta$ & $-1.28 \times 10^{-2}$ & $-0.094 \cdots +0.087$ \\
& $\overline{\Delta}$ & $+5.56 \times 10^{-2}$ & $0 \cdots 0.115$
\end{tabular}
\end{center}
\end{table}

\end{document}